\documentclass[12pt]{iopart}

\usepackage{iopams}
\usepackage{graphicx}
\usepackage{multirow}

\newcommand{\etc}{\textit{etc.}}
 \newcommand{\ie}{\textit{i.e.\ }}
 \newcommand{\eg}{\textit{e.g.\ }}
 \newcommand{\vs}{\textit{vs.}}
 
 \newcommand{\Eq}[1]{Eq.\ (\ref{#1})}

 \newcommand{\Fig}[1]{Figure \ref{#1}}

 \newcommand{\Kagome}{kagom\'{e}\ }

\begin{document}

\title{Site percolation on lattices with low average coordination numbers}

\author{Ted Y. Yoo$^{\dagger, 1}$, Jonathan Tran$^{\dagger, 1}$, Shane P. Stahlheber$^{\dagger, 1}$,
Carina E. Kaainoa$^{\dagger \dagger, 2}$, Kevin Djepang$^{\dagger
\dagger, 2}$, Alexander R. Small$^1$}


\address{$^1$Department of Physics and Astronomy, California State
Polytechnic University, Pomona, CA 91768 \\
$^2$Citrus College, Glendora, CA 91741\\ $^\dagger$ These authors
contributed equally, $^\dagger \dagger$ These authors contributed
equally} \ead{arsmall@csupomona.edu}

\begin{abstract}
We present a study of site and bond percolation on periodic lattices
with (on average) fewer than three nearest neighbors per site.  We
have studied this issue in two contexts: By simulating oxides with a
mixture of 2-coordinated and higher-coordinated sites, and by
mapping site-bond percolation results onto a site model with mixed
coordination number. Our results show that a conjectured power-law
relationship between coordination number and site percolation
threshold holds approximately if the coordination number is defined
as the average number of connections available between
high-coordinated sites, and suggest that the conjectured power-law
relationship reflects a real phenomenon requiring further study. The
solution may be to modify the power-law relationship to be an
implicit formula for percolation threshold, one that takes into
account aspects of the lattice beyond spatial dimension and average
coordination number.
\end{abstract}

\pacs{64.60.ah,  05.10.-a}
\noindent{\it Keywords}: percolation, lattices

\submitto{JSTAT}
\maketitle

\section{Introduction}

Percolation is a simple phase transition that can occur in a great
many systems that exhibit branched network
structure\cite{Stauffer1994, Sahimi1994}.  The basic idea is that
one occupies, at random, either the components of the system (often
sites on a lattice) or the links between the components (bonds
between sites). As the occupation probability per site (or bond)
increases, clusters of connected elements form and grow. When the
occupation probability exceeds a critical value $p_c$, there is a
cluster that spans the length of the system (or wraps, in
simulations with convenient periodic boundary conditions), from one
side to the other. In the context of a lattice, one could envision
the percolating cluster as being a continuous object that touches
both sides of the system. Percolation theory has applications in a
great many contexts, ranging from materials science topics such as
gelation or transport in porous media, to societally relevant
contexts such as disease propagation and forest fires.

\begin{figure}
    \includegraphics[scale=0.5]{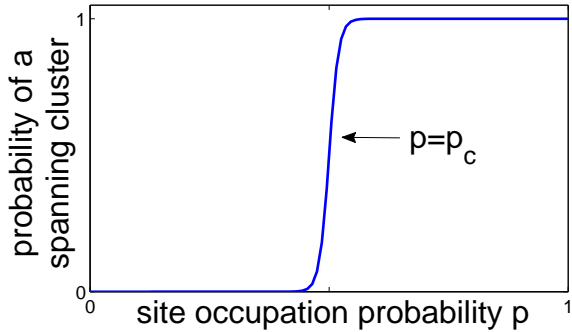}
    \caption{Example plot of percolation probability \vs\ site occupation probability.}
    \label{fig:RLschematic}
\end{figure}

The percolation threshold is only known exactly for special classes
of lattices, all in 2D\cite{Sykes1964,Ziff2006}. For most systems
one must use Monte Carlo simulations to compute
$p_c$\cite{Newman2001}. One determines $p_c$ in simulations by
occupying sites (or bonds) one-at-a-time in a random order, stopping
when the occupation of one more site (or bond) results in the
formation of a wrapping cluster. Different random orders give
percolation at different site occupation fractions, and after a
large number of lattices have been simulated one obtains a sigmoidal
plot of percolation probability (\ie probability of a wrapping
cluster forming) \vs\ site occupation probability
(\Fig{fig:RLschematic}). The percolation threshold is the point on
the plot at which the wrapping probability rises rapidly from zero
to one.

Despite the paucity of exact results, analytical formulas have been
proposed to provide approximate predictions (of varying accuracy)
for $p_c$ on different lattices, \eg \cite{Galam1996, Galam1997,
Suding1999, Wierman2005, Neher2008, Ziff2009}. One physically
intuitive approximation for the percolation threshold is a power-law
relation, noted by Galam and Mauger, between coordination number $z$
and percolation threshold $p_c$\cite{Galam1996}. Power-law behavior
is ubiquitous in critical phenomena, and the power-law relation is
known to give reasonably accurate predictions of $p_c$ for many
lattices with $z\geq 4$ in 2, 3, and sometimes higher dimensions.
Recently, it has also been shown to give fairly accurate predictions
of $p_c$ for $z$ as low as 3 in 3D\cite{Tran2013}.

The power-law relation has been shown to be problematic in regards
to its predictive power and
accuracy\cite{VanderMarck1997,VanderMarck1997b,VanderMarck1998,Babalievski1999,Wierman2005}.
For instance, lattices are grouped by Galam and Mauger into 3
different universality classes, but the theory provides no criterion
for determining which class a lattice should belong to.  Moreover,
there are many situations in which two lattices have the same
coordination number and dimensionality, but due to differences in
higher-order aspects of structure their percolation thresholds
differ, \eg the triangular and octagonal lattices in 2D (which have
the same site percolation threshold but different bond percolation
thresholds) \cite{VanderMarck1997b}, or the body-centered cubic and
stacked triangular lattices in 3D\cite{VanderMarck1997}. The
power-law scaling formula also lacks a number of other properties
that would be desirable in any accurate and widely-applicable
predictor of percolation thresholds\cite{Wierman2005}.  Despite
these problems, the power-law formula of Galam and Mauger remains of
some interest because it shows a common (if not universal) trend of
approximately power-law behavior in a phase transition.

It is natural to ask whether this power-law scaling of $p_c$
continues to hold for $z<3$. While it is impossible to construct an
interesting lattice in which \textit{all} of the sites have $z<3$
(it would just be a linear chain), one can still construct very
interesting lattices in which \emph{some} of the sites have
$z=2$\footnote{One could also consider a lattice in which some sites
have $z=1$, \ie some dangling atoms, analogous to the role of
singly-bonded hydrogen in a macromolecule, but from a percolation
standpoint such lattices are uninteresting. The occupation of a
$z=1$ site cannot connect two clusters to form a single cluster that
wraps around the system, and thus it can be eliminated from the
model without affecting any conclusions concerning connectivity and
transport.}, and others have $z\geq 3$. The $z=2$ sites can make the
lattice analogous to some oxides, with 2-coordinated sites
(analogous to oxygen atoms) joining together exactly 2 other sites.

Unfortunately, the conjectured power-law relation cannot remain
valid when the average coordination number gets sufficiently close
to 2. The power-law formula is:
\begin{equation}
    p_c = p_0 \left((d-1)(z-1) \right)^{-a}
    \label{eq:powerlaw}
\end{equation}
where $a= 0.6160$ and $p_0=1.2868$ for site percolation on 2D
Kagom\'{e} lattices and regular lattices in 3 or more dimensions,
and $a=0.3601$ and $p_0=0.8889$ for simple 2D lattices (\eg square,
honeycomb, and triangular). For $z=2$ and $d=3$ it predicts $p_c
<1$, when it should obviously be $1$ for $z=2$ (which corresponds to
a 1D chain). Given that this relationship cannot remain valid for
$z$ arbitrarily close to 2, but that it does work (to within $4\%$)
for $z=3$\cite{Tran2013}, it is reasonable to explore other low-$z$
cases to determine the accuracy of the power-law scaling relation.

The introduction of 2-coordinated sites along the bonds of a lattice
is equivalent to introducing a mixed site-bond problem, in which
sites are occupied with probability $p_s$ and bonds are occupied
with probability $p_b$\cite{Tarasevich1999}.  If we place a single
oxygen atom (2-coordinated site) along each bond, and treat the
2-coordinated sites the same as the other sites (\ie occupy them
with the same probability as other sites), we have a site-bond
problem with $p_b = p_s$. Physically, one could visualize this
problem as corresponding to an oxide, in which the two-coordinated
sites correspond to oxygen atoms and the higher-coordinated sites
correspond to atoms of higher valence. Alternately, if one wished to
visualize these problems in terms of communications networks or
similar phenomena, the 2-coordinated sites would correspond to
possible points of failure in connections between network nodes.  If
we introduce additional 2-coordinated sites (\ie more than one per
bond) and continue to treat them on an even footing with the other
sites we have a problem with $p_b<p_s$. This equivalence between a
problem with low coordination number and a site-bond percolation
problem enables us to use existing results for site-bond percolation
to probe site percolation on lattices with $z<3$.

However, setting up a lattice with average coordination number less
than 3 and finding its percolation threshold is not sufficient for
understanding the behavior of $p_c$ at low $z$. We also need a
proper quantitative measure of average coordination number on that
lattice. There are a number of plausible approaches that one could
take. Intuitively, one could compute the average number of atoms per
site, an approach based on the structure of the lattice.
Alternately, one could compute the average number of available bonds
between high-coordinated sites, an approach based on the
connectivity of the lattice at a particular occupation fraction, and
one that we will show to be useful in this work.

\begin{figure}
    \includegraphics{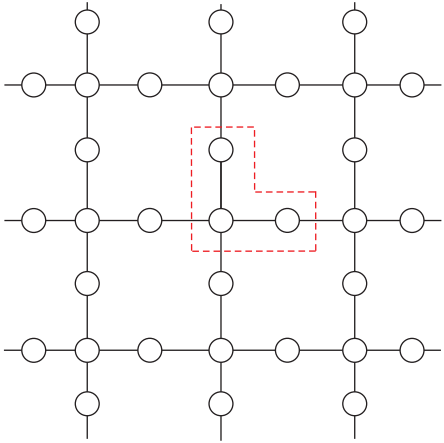}
    \caption{A simple example of a lattice with low average coordination number.  The underlying
    Bravais lattice is square, and the basis (outlined in red) is 3 atoms (one 4-coordinated, the others 2-coordinated).}
    \label{fig:Zeff}
\end{figure}

To illustrate the difference between these measures, consider the
lattice in \Fig{fig:Zeff}.  It is a square oxide structure, with a
3-atom basis outlined in red. The average coordination number per
site is easy to compute.  There are 2 atoms with 2 neighbors apiece,
and there is 1 atom with 4 neighbors. The average coordination
number is thus $\bar{z}=(1\cdot 4 + 2\cdot 2)/3\approx 2.667$.
However, forming a cluster that wraps around the system requires
linking 4-coordinated sites to one another. If we occupy sites with
probability $p$, and treat the 4-coordinated sites on the same basis
as 2-coordinated sites (\ie occupy each type of site with the same
probability) then the average number of sites occupied between the
4-coordinated sites is $4p$, and thus each 4-coordinated site has
connections to (on average) $4p$ other sites (which may or may not
be occupied). More generally, if we start with a lattice on which
each site has coordination number $z_l \geq 3$, introduce (possibly
multiple) 2-coordinated sites along some or all of the bonds, and
occupy those sites in such a way that there is a probability $p_b$
of having an unbroken bond between the $z_l$-coordinated sites, the
average number of accessible neighbors per $z_l$-coordinated site
will be $p_b z_l$. We will show here that choosing the later measure
(number of bonds between high-coordinated sites) leads to a more
robust scaling between percolation threshold and coordination number
for lattices with mixed coordination numbers.

In what follows we will study lattices with low coordination number
in two different ways: First, we will use Monte Carlo simulations to
compute site percolation thresholds on three oxide-type lattices
with average coordination number less than 3: the SiO$_2$ lattice (a
diamond lattice with 2-coordinated oxygen atoms along the bonds) and
two 3-coordinated lattices ((10,3)-a and
(10,3)-b\cite{Tran2013,Wells1977}) with 2-coordinated oxygen atoms
introduced along their bonds.  As a check on the scaling behavior of
$p_c$ at low coordination number we will also study a ``cubic oxide"
with average coordination number 3. We will show that the
percolation thresholds of these lattices are consistent (to better
than $2\%$, and in 2 cases to better than $0.1\%$) with the power
law conjectured by Galam and Mauger if average coordination number
is defined in terms of the average number of bonds between
high-coordinated sites rather than the average number of neighbors
per site. Second, using existing data on site-bond percolation
problems\cite{Tarasevich1999}, we will map mixed site-bond problems
onto pure site percolation problems with low coordination numbers
and show that power-law scaling of $p_c$ with $z$ again holds if the
average coordination number is measured by the average number of
bonds per high-coordinated site. Based on these results, we
conjecture that power-law scaling between $p_c$ and $z$ holds if one
defines $z$ in terms of available links between higher-coordinated
sites, leading to an implicit formula for $p_c$ that incorporates
aspects of lattice structure beyond dimensionality and average
coordination number.

\section{The oxide lattices studied}


We begin by briefly describing the four lattices that we are
studying here: silicon dioxide, (10,3)-a oxide, (10,3)-b oxide, and
cubic oxide.   We will describe the structure of these lattices in
their most symmetric forms, so that the reader can visualize them in
plausible physical realizations.  However, as in our previous
work\cite{Tran2013}, for computational purposes we deformed the bond
orientations so that the lattices could be mapped onto cubic grids,
albeit with fewer than six bonds to each site. Percolation
thresholds are governed only by the presence of the bonds, not their
orientations. Deforming the lattice in a way that preserves the
connections between sites does not change its percolation threshold,
but leads to greater convenience in enumerating sites. The process
of this deformation for (10,3)-a and (10,3)-b can be seen in greater
detail in our previous work.

\subsection{Silicon dioxide}

The silicon dioxide lattice is based on the diamond lattice, which
is a face centered cubic lattice whose primitive cell consists of
two lattice points. Each site is connected to four others in a
tetrahedral pattern\cite{Kittel2004}. The silicon dioxide lattice is
a diamond lattice in which each bond between tetrahedrally
coordinated sites has been replaced by a 2-coordinated site.  The
average coordination number per site on this lattice is
$(4+2+2)/3=2.67$.

\subsection{(10,3)-a oxide}

The (10,3)-a lattice is part of a larger family of 3-coordinated
lattices.  The lattices in this family go by a number of names, but
perhaps the most exhaustive exploration of this family of lattices
was by Wells, who coined the name (10,3)\cite{Wells1977}.  The 3
reflects the 3-fold coordination of the sites.  The number 10
reflects the fact that if one were to make a circuit on the lattice,
traveling site-by-site and visiting no site more than once before
the return to the starting point, the shortest path would be a
10-gon.  The -a reflects the fact that there are 7 lattices in this
family, named alphabetically: (10,3)-a, (10,3)-b, \etc\ In its most
symmetric realization, the (10,3)-a lattice is a bcc lattice with a
4-atom basis, each site having three neighbors. This lattice is of
interest for a number of reasons, including the fact that there are
real materials with this structure (\eg block
copolymers\cite{Cochran2004, Bates2005,Meuler2009}, molecular
magnets\cite{Liu2004,Gil2010}, and even butterfly
wings\cite{Saba2011}).  Additionally, it has unusually high
symmetry, possessing a property known as ``strong
isotropy"\cite{Sunada2008}, shared with only one other 3D lattice
(diamond).  The structure of (10,3)-a is illustrated in our previous
paper on this subject\cite{Tran2013}.

\begin{figure}
    \includegraphics[scale=0.5]{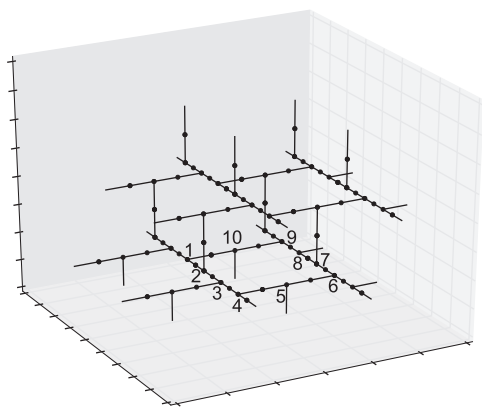}
    \caption{The (10,3)-a oxide lattice, with bond angles deformed to fit onto a cubic grid for convenience.
    We number ten of the 3-coordinated sites (but not the 2-coordinated sites joining them) to show that the shortest
    circuit along the underlying (10,3)-a lattice is a 10-gon.}
    \label{fig:103aox}
\end{figure}

The oxide form that we will study for this lattice is one in which a
2-coordinated site is inserted between each 3-coordinated site. If
we denote the 2-coordinated sites as O (for oxygen) and the
3-coordinated sites as X, the stoichiometric formula of this
structure is X$_2$O$_3$, but it is important to note that this is
\textit{not} the typical structure of a Group III oxide (\eg the
most common form of aluminum oxide has a coordination number higher
than 3\cite{Ishizawa1980}). For our purposes here, (10,3)-a oxide is
of interest for being a convenient lattice with a low coordination
number. For each pair of 3-coordinated sites there are 3
2-coordinated sites, giving an average coordination number of
$(2\cdot 3+3\cdot 2)/5=2.4$.  We illustrate it \Fig{fig:103aox},
with bond orientations deformed to fit the lattice onto a cubic grid
for convenience in enumerating sites.

\subsection{(10-3)-b oxide}

\begin{figure}
    \includegraphics[scale=0.5]{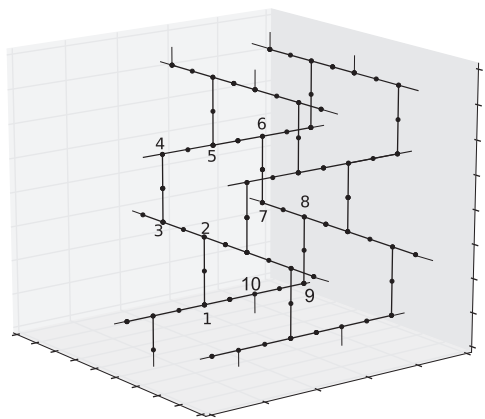}
    \caption{The (10,3)-b oxide lattice, with bond angles deformed to fit onto a cubic grid for convenience.
    We number ten of the 3-coordinated sites (but not the 2-coordinated sites joining them) to show that the shortest
    circuit along the underlying (10,3)-b lattice is a 10-gon.}
    \label{fig:103box}
\end{figure}

The (10,3)-b lattice is similar to (10,3)-a, in that the shortest
circuit that returns to a site is a 10-gon, and that the primitive
cell has four atoms. The effective coordination number of the oxide
is again 2.4. However, the lattice contains an extra structural
degree of freedom. Even if all bond angles are 120 degrees, and all
bonds are equal length (\ie a highly symmetric realization of the
lattice), we can continuously deform the lattice in a manner that
uniformly changes the spacing between planes without altering any
bond lengths or angles.  In the maximally symmetric case, the
lattice is body-centered tetragonal with a four-atom basis. If,
however, we compress the lattice to decrease the spacing between
planes, it begins to look like a 3D generalization of a honeycomb
lattice, being a stacking of interwoven honeycomb planes.  For this
reason, the (10,3)-b lattice has been referred to as the ``hyper
honeycomb" lattice, and has attracted some interest in the study of
spin systems\cite{Lee2014}.  We will consider an oxide form which,
again, has a 2-coordinated site along each bond.  We illustrate it
\Fig{fig:103box}, with bond orientations deformed to fit the lattice
onto a cubic grid for convenience in enumerating sites.

\subsection{Cubic oxide}

Cubic oxide is a simple cubic lattice (6-coordinated sites) with
2-coordinated oxygen sites along the bonds between the 6-coordinated
sites.  There are 6 oxygen sites bonded to each 6-coordinated site.
The oxygen sites are shared between 2 neighbors, so the
stoichiometric formula is XO$_3$ and the average coordination number
is $\bar{z}=(1\cdot 6+3\cdot 2)/4=3$. While we have studied the case
of $z=3$ in our previous work on 3-coordinated lattices, this case
was included to see if there are significant differences between a
homogeneous lattice (all sites the same coordination number) and an
oxide with a different underlying structure but the same average
coordination number.

\section{Computational methods}

\subsection{The Newman-Ziff algorithm}

We determined $p_c$ with the algorithm of Newman and
Ziff\cite{Newman2001}.  We described our implementation in previous
work\cite{Tran2013}.  In brief, the Newman-Ziff algorithm occupies
sites or bonds on a lattice in a random order.  When a new site (or
bond) is occupied, the program checks to see if it borders an
existing cluster of sites (or bonds).  If so, it joins that cluster.
If the new site (or bond) bridges two existing clusters, the
clusters are merged.  Finally, if the occupation of a new site (or
bond) joins two parts of an existing cluster and causes them to wrap
completely around, then percolation has occurred, and the program
moves on to randomly occupy sites or bonds on a new, fresh lattice.
By repeating this process $N_L$ times, one can get the fraction
$R_L$ of lattices of linear dimension $L$ (\ie number of unit cells
in the system is $L^3$) that wrap when a given number $n$ of the
sites are occupied.

In order to obtain wrapping probability $R_L$ as a function of site
occupation \textit{probability} $p$, one convolves $R_L(n)$ with the
binomial distribution:
\begin{eqnarray}
    R_L(p) = \sum_{n=0}^N \left( \begin{array}{c} N \\ n \end{array}
    \right) p^n (1-p)^{N-n} R_L(n)
    \label{eq:binomial}
\end{eqnarray}
The convolution amounts to a weighted sum over all possible
realizations of a lattice in which sites are occupied with
probability $p$. For each possible occupation number we multiply the
probability of that occupation number occurring (from the binomial
distribution) by the wrapping probability for that occupation
number.

The plot of $R_L(p)$ has a steep rise at the percolation threshold
$p_c$, and one could obtain a reasonable estimate of $p_c$ simply by
looking for the point on the graph with the steepest slope. However,
one can obtain a higher precision estimate by comparing plots of
$R_L(p)$ for several different values of the linear dimension $L$.
The percolation threshold is the point at which the wrapping
probability crosses over from 0 to 1, and the width of this
cross-over region gets smaller as $L$ increases.  Consequently,
above $p_c$ the wrapping probability increases as $L$ increases, and
below $p_c$ the wrapping probability decreases as $L$ increases.  If
we make a plot of $R_L$ \vs\ $L$ for different site occupation
probabilities, this plot will be flattest at $p=p_c$.  This method
has been used to determine $p_c$ by us and others in previous
work\cite{Tran2013,Martins2003}.

\subsection{The uncertainty in the percolation threshold}

There is a straightforward way to determine the uncertainty in an
estimate of $p_c$.  If one repeatedly generates $N_L$ lattices with
the same fraction of occupied sites, one would expect the standard
deviation of the fraction that wraps to be given by the binomial
distribution: $\sqrt{R_L(1-R_L)/N_L}$. This is the magnitude of
vertical fluctuations in the $R_L$ \vs\ $p$ graph, approximately
$3\%$ for simulations of $N_L=10^3$ lattices wrapping with
probability $R_L\approx 0.5$ (the numbers used here).  The figures
below do indeed exhibit vertical fluctuations of that order.

To go from vertical fluctuations to horizontal fluctuations (\ie the
effect on the estimate of $p_c$), one must divide by the slope of
the $R_L$ \vs\ $p$ graph, giving:
\begin{equation}
    \delta p_c = \sqrt{\frac{R_L(1-R_L)}{N_L}}\left/\frac{dR_L}{dp}
    \right.
    \label{eq:uncert}
\end{equation}
By way of comparison, when we generated $N_L=10^3$ lattices of size
$L^3=128 \times 128 \times 128$ unit cells, typically with 4 or more
sites (depending on the lattice) per unit cell, this amounted to
producing of order $10^{10}$ random numbers, which should give
uncertainties of order $10^{-5}$ or larger.  For the cases shown
below, uncertainties were estimated to be between $2\cdot 10^{-5}$
and $5\cdot 10^{-5}$, consistent with the number of random numbers
generated.

We convolved our $R_L(n)$ data with the binomial distribution twice.
On the first pass, we worked with a coarse-grained distribution,
varying the site occupation probability $p$ in steps of
$1/$(smallest lattice size).  We looked for the approximate
intersection of the $R_L$ \vs\ $p$ curves to get an approximate
$p_c$, and used the values of $R_L$ and $dR_L/dp$ to get an
approximate uncertainty in $p_c$.  After that, we again convolved
$R_L(n)$ with the binomial distribution, this time varying $p$ in
steps of the estimated uncertainty.  After the second convolution,
we looked for the value of $p$ that made $R_L(p)$ flattest (\ie
smallest slope in a least squares fit) as a function of $L$, to get
a better estimate of $p_c$ and its uncertainty.

Using this method for determining $p_c$ and the uncertainty in
$p_c$, we previously\cite{Tran2013} determined the site percolation
threshold of the 3D simple cubic lattice to be in close agreement
with the literature value\cite{Lorenz1998a}.  As a further check on
our work, before computing the percolation threshold of silicon
dioxide we used simulations to determine the site and bond
percolation thresholds of the diamond lattice, getting results in
good agreement with the most precise available literature
values\cite{Xu2014}. Besides confirming the reliability of our
implementation of the Newman-Ziff algorithm, this also gives us
confidence that our code correctly represents the silicon dioxide
lattice, as the code for silicon dioxide was based on the test code
for the diamond lattice (due to the close relationship between the
lattices).


\section{Computational results for site percolation thresholds of oxides}

\begin{figure}
    \includegraphics[scale=0.5]{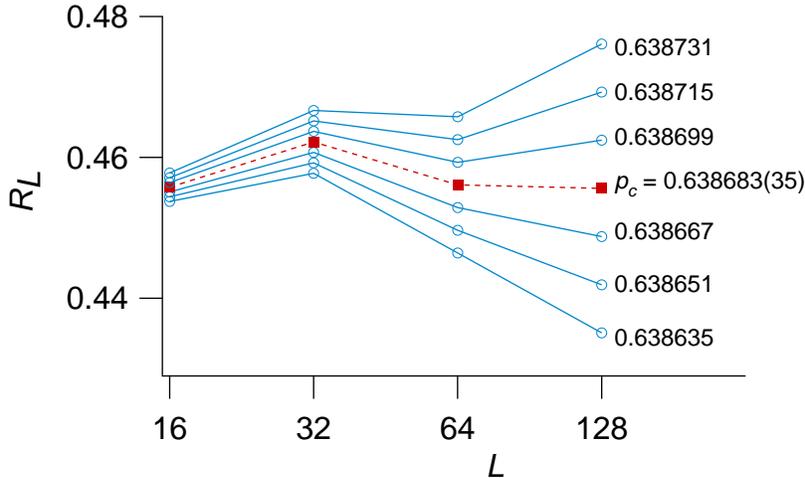}
    \caption{Wrapping probability $R_L$ \vs\ $L$ for site percolation on
    the silicon dioxide lattice, for different
    occupation probabilities $p$. The $p$ that produces the flattest
    overall trend (dashed red line) is taken to be the percolation
    threshold $p_c$.}
    \label{fig:RLSiO2}
\end{figure}

\begin{figure}
    \includegraphics[scale=0.5]{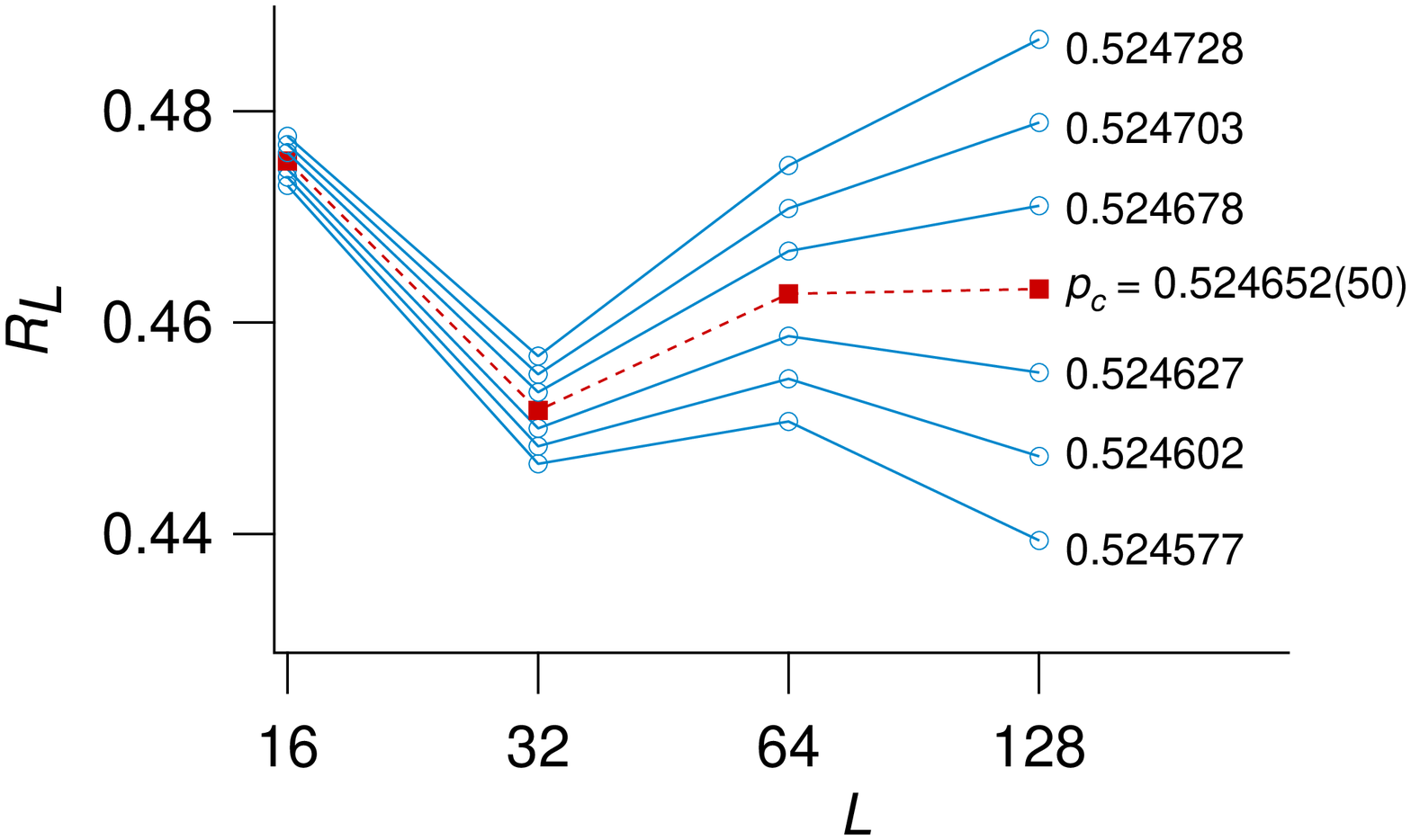}
    \caption{Wrapping probability $R_L$ \vs\ $L$ for site percolation on
    the cubic oxide lattice, for different
    occupation probabilities $p$. The $p$ that produces the flattest
    overall trend (dashed red line) is taken to be the percolation
    threshold $p_c$.}
    \label{fig:RLcubeox}
\end{figure}

\begin{figure}
    \includegraphics[scale=0.5]{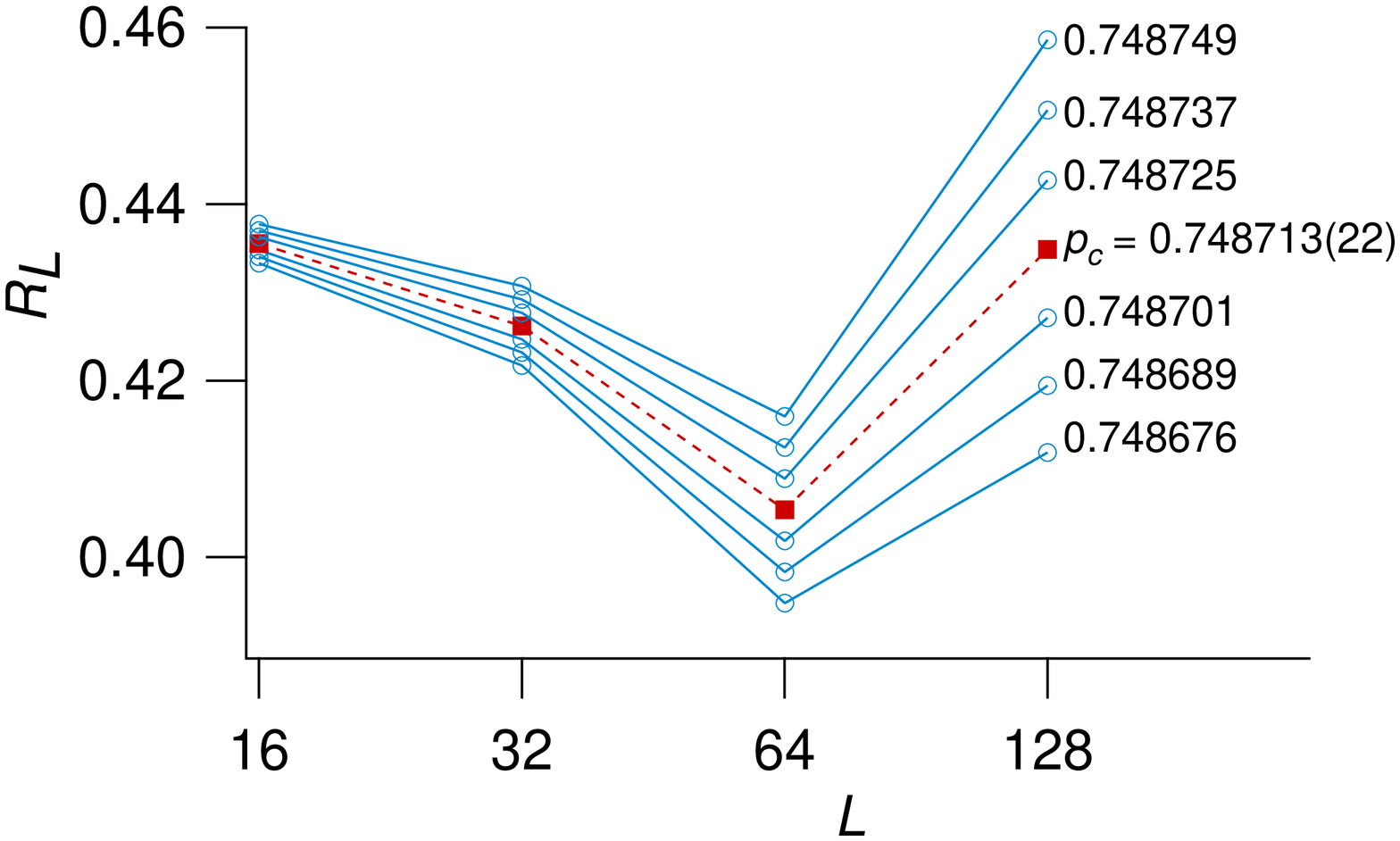}
    \caption{Wrapping probability $R_L$ \vs\ $L$ for site percolation on
    the (10,3)-a oxide lattice, for different
    occupation probabilities $p$. The $p$ that produces the flattest
    overall trend (dashed red line) is taken to be the percolation
    threshold $p_c$.}
    \label{fig:RL103a}
\end{figure}

\begin{figure}
    \includegraphics[scale=0.5]{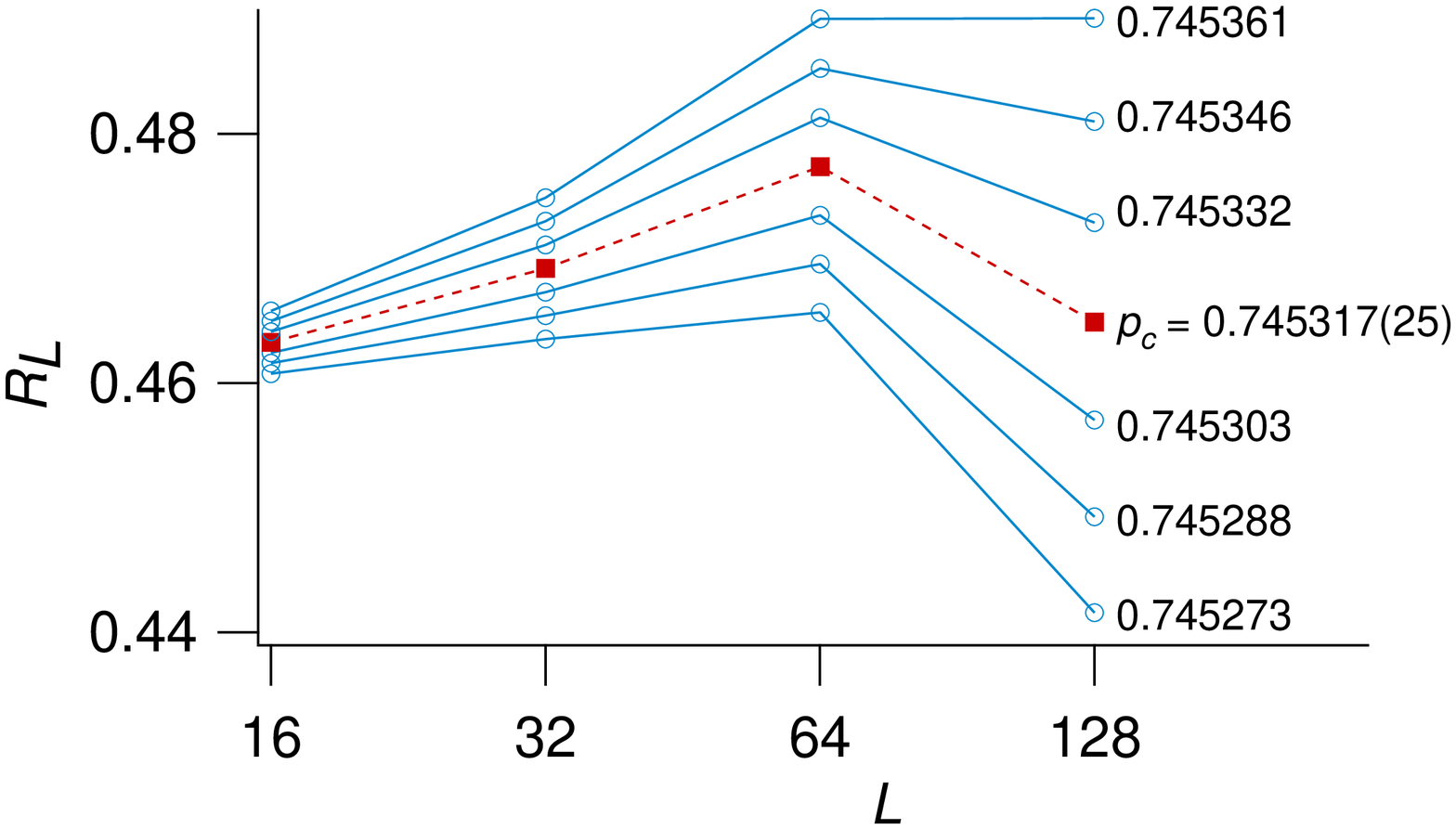}
    \caption{Wrapping probability $R_L$ \vs\ $L$ for site percolation on
    the (10,3)-b oxide lattice, for different
    occupation probabilities $p$. The $p$ that produces the flattest
    overall trend (dashed red line) is taken to be the percolation
    threshold $p_c$.}
    \label{fig:RL103b}
\end{figure}

Figures \ref{fig:RLSiO2} through \ref{fig:RL103b} show $R_L$ \vs\
$L$ for silicon dioxide, cubic oxide, (10,3)-a oxide, and (10,3)-b
oxide. For each type of lattice we generated $N_L=10^3$ cases.  Each
figure shows plots for selected values of $p$, spaced by intervals
of the uncertainty (as computed from \Eq{eq:uncert}). There are
fluctuations in the graphs of $R_L$ \vs\ $L$, but they are of order
$\sqrt{R_L(1-R_L)/N_L}$ as discussed above, and the mean values
($\approx 0.45$) of the wrapping probability $R_L$ at the
percolation threshold are very close to values found at $p_c$ in
other investigations of 3D site percolation\cite{Martins2003}.  For
a given value of $L$, all values of $R_L$ fluctuate in the same
direction and by approximately the same amount (irrespective of the
value of $p$) because they are derived from the same set of
simulated lattices. The percolation thresholds of these oxides are
summarized in Table~\ref{tab:table1}.

\begin{table}
\caption{\label{tab:table1}Site percolation thresholds for oxides
studied here. Uncertainties in $p_c$ estimates are given in
parentheses, and refer to the last two digits. }

\begin{indented}
\lineup
\item[]\begin{tabular}{ccll}
\br lattice & average neighbors per site & $p_c z_l$ & $p_c$ (site)  \\
 \hline
 (10,3)-a oxide & $2.4$ & $2.246$ & $0.748713(22)$  \\
 (10,3)-b oxide & $2.4$ & $2.236$ & $0.745317(25)$  \\
 Silicon dioxide & $2.66...$ & $2.555$ & $0.638683(35)$ \\
 Cubic oxide & 3 & $3.148$ & $0.524652(50)$ \\
\br
\end{tabular}
\end{indented}
\end{table}

\begin{figure}
    \includegraphics[scale=0.5]{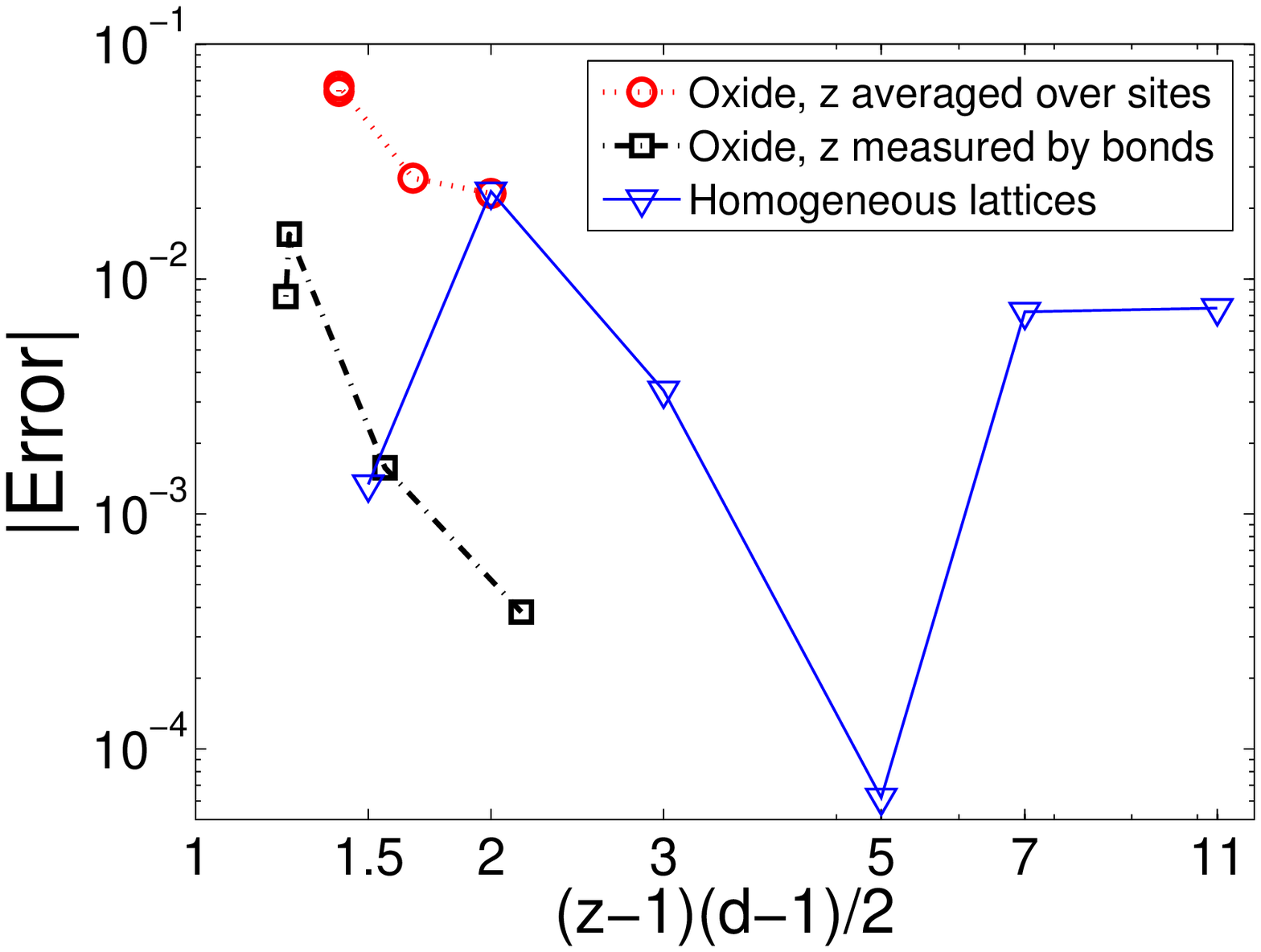}
    \caption{Absolute value of the error in the prediction of the
    approximate power-law formula, as a function of coordination number or (for oxides)
    average coordination number. The oxides are (\emph{from left to right})
    (10,3)-b oxide, (10,3)-a oxide, SiO$_2$, and cubic oxide. The six regular
    lattices are (\emph{from left to right}) 2D \Kagome\cite{Ziff1997},
    (10,3)-a\cite{Tran2013}, diamond\cite{Xu2014}, simple cubic\cite{Lorenz1998a}, body-centered
    cubic\cite{Lorenz1998a}, and face-centered cubic\cite{Lorenz1998}. The factor of $(d-1)/2$ is
    included in the horizontal axis to enable comparisons between 3D lattices and the
    2D \Kagome lattice.}
    \label{fig:error}
\end{figure}

\begin{figure}
    \includegraphics[scale=0.5]{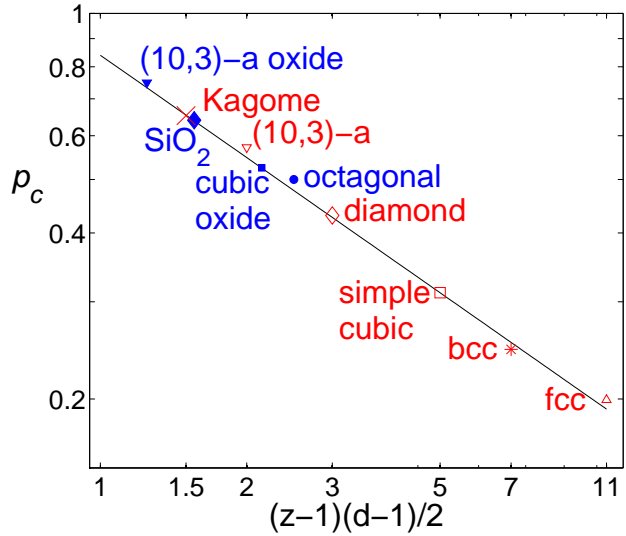}
    \caption{Log-log plot comparing percolation thresholds of different lattices.  Lattices with uniform coordination numbers
    are shown in red, and mixed coordination numbers are shown in blue. We include here the 2D
    octagonal lattice (discussed in section \ref{sec:Discussion}).  The black line shows the power-law scaling relation from \Eq{eq:powerlaw},
    with $p_0$ and $a$ from the second universality class identified by Galam and Mauger.
    The horizontal scale includes a factor of $(d-1)/2$
    to enable comparisons between 3D lattices and the 2D \Kagome and octagonal lattices.  For lattices with variable numbers
    of nearest neighbors, $z$ is the average number of connections between high-coordinated sites at $p_c$.}
    \label{fig:sitescaling}
\end{figure}

With these numbers in hand, we can compare the computed percolation
thresholds with the predictions of power-law scaling. As discussed
above, we have at least two plausible methods for quantifying the
average coordination number on a lattice, and each measure, when
used in the power-law, will give a different prediction for the site
percolation threshold. In \Fig{fig:error}, we show the errors
produced by each method for the four oxides studied, as well as the
octagonal lattice (which has mixed coordination number and is
discussed below in section \ref{sec:Discussion}). For comparison, we
also show the errors for several well-known uniform lattices (all
sites with same coordination number). Quantifying average
coordination number according to the average number of bonds
available to a high-coordinated site improves the predictions by at
least an an order of magnitude for each oxide. When we then plot
$p_c$ as a function of $z-1$ we see that all of the oxides are very
close to the same line as other lattices in \Fig{fig:sitescaling}.

Interestingly, the errors for the oxides are actually comparable to
or smaller than the errors for most of the regular,
higher-coordinated lattices.  The errors for cubic oxide and silicon
dioxide are $\approx 3\cdot 10^{-4}$ and $\approx 1.5\cdot 10^{-3}$
respectively, smaller than the errors that Galam and Mauger obtained
when they tried to derive effective coordination numbers for uniform
3D lattices\cite{Galam1997}. The results are less impressive for the
(10,3) oxides, but are still a definite improvement (order of
magnitude reduction in error) over the results obtained by using an
arithmetic average coordination number.  Even for the lowest-$z$
oxide, errors are of the same order of magnitude as the errors
obtained by Galam and Mauger for uniform 3D lattices of higher
coordination number. The fact that the formula of Galam and Mauger
happens to be accurate for some lattices is not surprising; what is
more interesting is that the formula remains reasonably accurate
when one applies it to site-bond problems. The good agreement with
power-law scaling for an effective $z$ as low as 2.246 is especially
significant in light of the fact that for the higher-coordinated
uniform lattices the coordination number is unambiguously defined,
whereas there are multiple plausible definitions for the average
coordination number of an oxide.

\section{Site-bond percolation as a site percolation problem with low coordination number}

We do not have to restrict our attention to oxides if we want to
study low coordination numbers.  We can also study site-bond
percolation problems, in which (as discussed above) sites are
occupied with probability $p_s$ and bonds with probability $p_b$.
One can hold the bond occupation probability fixed and vary the site
occupation probability (or vice-versa) to see when a wrapping
cluster forms, and thereby determine the site percolation threshold
as a function of bond occupation probability $p_b$ (or vice-versa).
Removing some of the bonds reduces the average coordination number
from $z_l$ (the coordination number of the underlying lattice) to
$p_b z_l$.

\begin{figure}
    \includegraphics[scale=0.55]{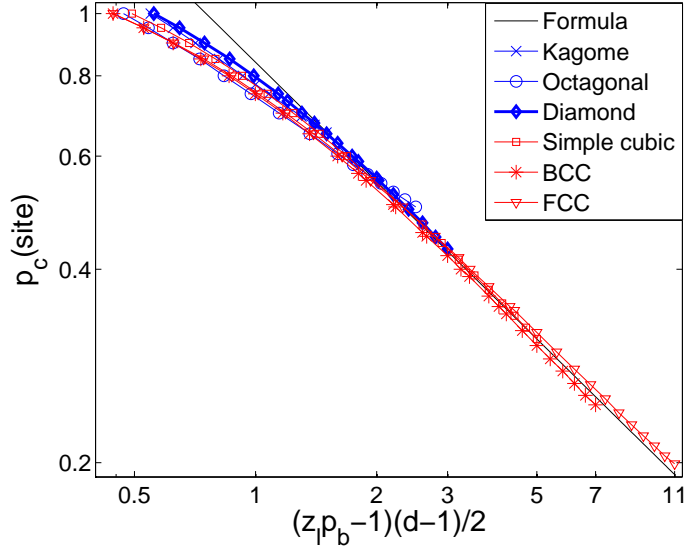}
    \caption{Site percolation threshold (in site-bond percolation) as a function of mean number of available bonds $z_l p_b$ and dimension $d$
    for the universality class of 3D lattices (and also 2D Kagom\'{e} lattices) identified by Galam and Mauger.  We
    include the dimension on the horizontal axis because the formula depends on $(z-1)(d-1)$ and the Kagom\'{e} lattice
    is 2D. Results for the diamond, octagonal, and \Kagome lattices are in blue to highlight lattices with low coordination number.
    The black line shows the power-law scaling relation from \Eq{eq:powerlaw},
    with $p_0$ and $a$ from the second universality class identified by Galam and Mauger.}
    \label{fig:3Dzeff}
\end{figure}

\begin{figure}
    \includegraphics[scale=0.45]{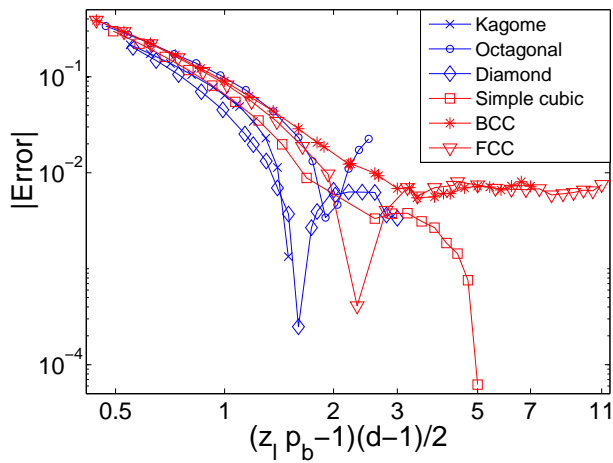}
    \caption{Difference between prediction of power-law scaling and site percolation threshold as a function
    of average coordination number in site-bond calculations.}
    \label{fig:sitebonderrors}
\end{figure}

In \Fig{fig:3Dzeff} we show plots of $p_c$ \vs\ $p_b z_l$, using
data from Tarasevich and van der Marck\cite{Tarasevich1999}. We
considered several lattices that Galam and Mauger conjectured to
fall into the same universality class. The right end of each data
series lies close to the line representing the power law of Galam
and Mauger, as well as the 2D octagonal lattice (discussed below).
The agreement ultimately breaks down at sufficiently small effective
coordination numbers, because eventually the problem of constructing
a wrapping cluster is dominated by the low number of occupied bonds,
and scaling relations valid for site percolation no longer apply.
Nonetheless, for each lattice considered, the site percolation
threshold initially follows the power-law scaling relation as some
of the bonds are removed. Quantitative agreement is confirmed by
examining the errors in \Fig{fig:sitebonderrors}.  The mere fact
that the formula of Galam and Mauger works in these cases (which
have been conjectured to belong to a universality class) is less
interesting than the fact that the formula continues to work for
those same cases when we generalize from site percolation to
site-bond percolation.

\begin{figure}
    \includegraphics[scale=0.5]{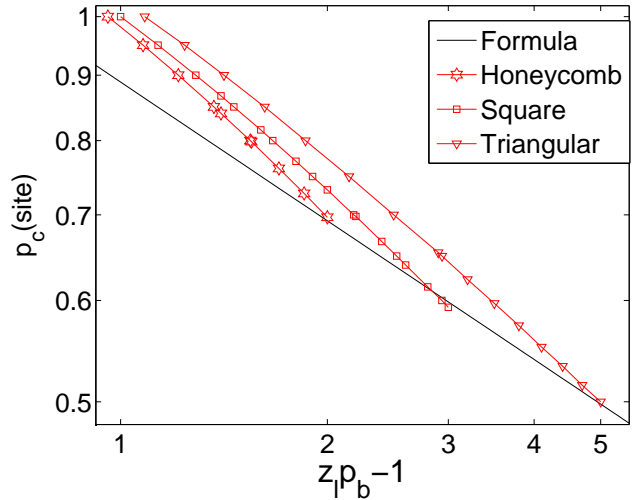}
    \caption{Site percolation threshold (in site-bond percolation) as a function of mean number of available bonds $z_l p_b$ and dimension $d$
    for the universality class of 2D honeycomb, square, and triangular lattices identified by Galam and Mauger.
    The black line shows the power-law scaling relation from \Eq{eq:powerlaw},
    with $p_0$ and $a$ from the first universality class identified by Galam and Mauger. }
    \label{fig:2D}
\end{figure}

However, when we performed the same analysis for the universality
class of 2D lattices identified by Galam and Mauger (honeycomb,
triangular, and square)\cite{Galam1996}, the results are not
well-approximated (even for $p_b \approx 1$) by the power-law
scaling of site percolation threshold.  We have no good explanation
for this phenomenon.  It cannot be a consequence of dimensionality,
as the 2D \Kagome lattice results follow the power-law scaling
behavior of regular lattices.  While the concept of 2 distinct
universality classes at low dimension has been met with important
objections\cite{VanderMarck1997,Wierman2005}, the differences
between \Fig{fig:3Dzeff} and \Fig{fig:2D} show that lattices that
(approximately) follow a common scaling law for pure site and bond
percolation also (approximately) also follow the same scaling law
for mixed site-bond percolation.  While the universality classes
proposed by Galam and Mauger have significant defects (\eg no
criterion to predict which lattices should belong to which class),
this difference in scaling behavior suggests that there may be some
underlying phenomenon meriting further study.  It is worth noting
that Ziff and Gu have developed a useful and reasonably accurate
(better than $3\%$) approximate formula for site-bond percolation on
the honeycomb lattice\cite{Ziff2009}, but we are not aware of any
generalization that we can use for other 2D lattices.

\section{Discussion} \label{sec:Discussion}

The measures we considered here for average coordination number are
not the only possible approaches. Galam and Mauger proposed that one
could define an effective coordination number as the value of $z$
that minimizes the sum of the errors in the site and bond
percolation thresholds (with errors defined as the difference
between the actual percolation threshold and the predictions of
power-law scaling).  However, this is an after-the-fact definition
of effective coordination number, requiring knowledge of the true
percolation threshold.  One cannot derive this effective
coordination number from a direct examination of lattice geometry,
but only from already knowing the percolation threshold. Our
approach, in which we define $z_l p_c$ as the effective coordination
number at the percolation threshold, can be used to obtain a
prediction of $p_c$ without \textit{a priori} knowledge of the
percolation threshold. If one uses power-law scaling, one gets an
implicit formula for an oxide's site percolation threshold $p_c$:
\begin{equation}
    p_c = p_0 \left((d-1)(z_l p_c-1) \right)^{-a}
\end{equation}
This equation can be solved for $p_c$, as long as one knows $z_l$,
$d$, $p_0$, and $a$.

\begin{figure}
    \includegraphics[scale=0.35]{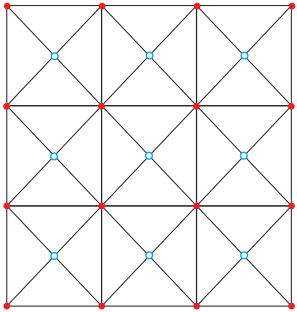}
    \caption{Octagonal lattice, often called the ``Union Jack" lattice,
    with a mix of 8-coordinated (solid red) and 4-coordinated (open blue) sites.}
    \label{fig:Octagonal}
\end{figure}

We can generalize this procedure. Consider a more complicated
lattice, such as the octagonal lattice (\Fig{fig:Octagonal}) with
site percolation threshold $p_c^{(s)}=0.5$ (because it is fully
triangulated) \cite{VanderMarck1997b}. We consider the octagonal
lattice because (1) its combination of site and
bond\cite{VanderMarck1997b} percolation thresholds make it a
plausible candidate for being in the same universality class as the
\Kagome lattice\cite{Galam1998} and (2) its structure of mixed
coordination numbers is more complex than the oxides considered
above, but is nonetheless simple to study.  Let us assume that
formation of a percolating cluster is dominated by the 8-coordinated
sites, and that the 4-coordinated sites primarily serve to
facilitate links between 8-coordinated sites. Each 8-coordinated
site has direct bonds to 4 other 8-coordinated sites, as well as 4
4-coordinated sites. Occupation of these other 4-coordinated sites
is not the only way to reach other 8-coordinated sites, but we can
begin by trying to ``average out" the 4-coordinated sites, modeling
this in a manner analogous a site-problem.

In our approximation, where we replace 4-coordinated sites with
bonds, at the percolation threshold $4p_c$ of the 4-coordinated
neighbors will be occupied and permit access to 8-coordinated
neighbors.  Additionally, there are direct bonds to 4 other
8-coordinated sites. The average coordination number is thus
$4+4p_c$. If we assume the power-law scaling relation to hold, we
get:
\begin{equation}
    p_c = p_0 \left( (d-1)(4+4p_c-1) \right)^{-a}
    \label{eq:octagonal}
\end{equation}
This is an implicit equation for $p_c$.  More importantly, it was
derived by taking into account features of the lattice beyond
nearest neighbors, addressing one of the common criticisms of the
Galam-Mauger power law.  If we solve \Eq{eq:octagonal} numerically
we get that the site percolation threshold is $0.4818$.  This is
reasonably close to the exact site percolation threshold of $1/2$,
but still not resounding agreement.  However, when we use this
prediction of the site percolation threshold to compute an effective
coordination number ($4+4p_c=5.93$) and then use that in Galam and
Mauger's formula for the bond percolation threshold, we get the
prediction $p_c^{(b)}=0.3247$, which agrees with the bond
percolation threshold determined from Monte Carlo simulations
($p_c^b =0.3237$) to $0.3\%$.  While our inclusion of the octagonal
lattice in the same universality class as \Kagome is just
conjecture, the site bond results for octagonal in \Fig{fig:3Dzeff}
follow the same trend as Kagome and the 3D lattices, suggesting that
its inclusion in this conjectured universality class is a hypothesis
worthy of further study.

Of course, our approach to ``averaging out" lower-coordinated sites
is \textit{ad hoc} and gives no obvious prescription for more
general lattices.  We only present it as the beginning of an idea
that might be generalized to more complicated lattices: Treating the
higher-coordinated sites as the key players in the formation of a
percolating cluster, and the lower-coordinated sites as providing an
average number of bonds between the higher-coordinated sites.  By
removing the lower-coordinated sites from explicit consideration,
one is removing the shortest length scales from the analysis,
bringing in ideas analogous to renormalization group calculations.
However, renormalization treatments of percolation do not assume
compliance with an empirical power law derived from Monte Carlo
simulation results, unlike our use of the Galam Mauger formula.
Nonetheless, power laws are closely related to the idea of scale
invariance, raising the question of whether our approach to oxides
and the octagonal lattice might have connections to renormalization
ideas.  However, given that there are demonstrated examples of
lattices for which our approach does \textit{not} support a
power-law relationship between site percolation threshold and
average coordination number, a more rigorous exploration of these
ideas would have to provide some criterion for identifying the
lattices that can be treated by this approach.  Criteria that relate
lattice topology to properties of percolating clusters seem
especially promising for further insight\cite{Neher2008}.

\section{Conclusions}

In conclusion, we have shown that a conjectured power-law
relationship between site percolation threshold and average
coordination number holds for a number of lattices with average
coordination number less than 3. These lattices are closely
analogous to oxide materials. The quantitative accuracy of the
power-law scaling relationship depends on how the average
coordination number is defined and measured, with better agreement
when we define average coordination number as the average number of
available bonds between higher-coordinated sites. When we apply that
definition of average coordination number to site-bond percolation
problems, we find that existing simulation results are roughly
consistent with the power-law scaling conjecture for site
percolation.  More interestingly, this approach turns the power-law
scaling formula into an implicit formula for $p_c$, one that can (in
some cases) take into account next-nearest neighbors. Finally,
although there are some key cases in which our approach fails to
give  good agreement between simulation results and the conjectured
power-law scaling behavior, the relevant lattices are those that had
previously been conjectured to fall into a distinct universality
class.  While there are still significant problems with the
power-law scaling formula and the conjectured universality classes,
our results suggest that a re-examination is merited, to see if
there are ways to refine these conjectures to take into account
additional features of the lattice structure and refine estimates of
percolation thresholds.

\ack This work was supported in part by the Citrus College Race to
STEM program, funded by the US Department of Education. Shane
Stahlheber was also supported by the Kellogg University Scholars
program.  The writing of this paper was completed on a sabbatical in
the Laboratory for Fluorescence Dynamics at UC Irvine.  We thank
Robert Ziff for many useful discussions.


\section{Bibliography}
\bibliographystyle{unsrt}

\end{document}